\documentclass[runningheads]{llncs}
\usepackage{graphicx}
\graphicspath{{./img/}}
\usepackage{amssymb}
\usepackage[T1]{fontenc}
\usepackage{graphicx}
\begin{document}

\title{On the Convergence of Malleability and the HPC PowerStack: Exploiting Dynamism in Over-Provisioned and Power-Constrained HPC Systems}
%
%

\titlerunning{On the Convergence of Malleability and the HPC PowerStack}
%
\author{Eishi Arima\inst{1} \and
Isa{\'i}as~A. Compr{\'e}s\inst{1} \and
Martin Schulz\inst{1,2}}
%

%
\institute{Technical University of Munich, Garching, Germany \and
Leibniz Supercomputing Centre, Garching, Germany\\
\email{\{eishi.arima,compresu,schulzm\}@in.tum.de}}
\maketitle              
\begin{abstract}
Recent High-Performance Computing (HPC) systems are suffering from severe problems, such as massive power consumption, while at the same time significantly under-utilized system resources. Given the power consumption trends, future systems will be deployed in an over-provisioned manner where more resources are installed than they can afford to power simultaneously. 
In such a scenario, maximizing resource utilization and energy efficiency, while keeping a given power constraint, is pivotal.
Driven by this observation, in this position paper we first highlight the recent trends of resource management techniques, with a particular focus on malleability support (i.e., dynamically scaling resource allocations/requirements for a job), co-scheduling (i.e., co-locating multiple jobs within a node), and power management. Second, we consider putting them together, assess their relationships/synergies, and discuss the functionality requirements in each software component for future over-provisioned and power-constrained HPC systems. Third, we briefly introduce our ongoing efforts on the integration of software tools, which will ultimately lead to the convergence of malleability and power management, as it is designed in the HPC PowerStack initiative.

\end{abstract}

\section{Introduction}
The power consumption of top-class supercomputers or High-Performance Computing (HPC) systems have been increasing considerably over the past few decades. 
As a result, one of the most powerful supercomputers in the world now consumes an enormous amount of power, almost hitting 30MW~\cite{top500}. 
Meanwhile, energy costs have been raising significantly in general, 
and thus setting a power constraint on the entire HPC system in order to keep within a budgetary upper limit is becoming more and more critical. 
As a consequence, future HPC systems will be deployed in an \textit{over-provisioned} manner, i.e., installing more resources than the facility can (or wants to) afford in terms of supplied power at one time, and will be operated under a certain power constraint depending on the operation cost at the time, and using techniques like active power shifting to direct the limited resource power to the system components that require it most to optimize performance and/or throughput. 

For this approach to work, though, we require significant flexibility in the entire system software stack.
One promising solution for this is supporting dynamic malleability, i.e., dynamically scaling resource request/allocation to exploit the dynamism inside of an application and the entire system. 
Because current standard resource schedulers in HPC employ static resource allocation policies, there is a significant room for system efficiency improvement by introducing dynamism at this level. 
Another promising solution is co-scheduling, i.e., co-locating multiple jobs that utilize complementary resources on the same node. 
As a compute node in an HPC system is becoming increasingly fat with heterogeneous processing elements, co-scheduling is indispensable to fully utilize the resources inside a node. 

In this position paper, we explicitly target the near-future over-provisioned and power-constrained HPC systems and consider applying these novel approaches, which both boil down to sophisticated resource handling mechanisms, to these systems. 
More specifically, we first highlight the trends of HPC architectures, malleability support, co-scheduling, and power-aware HPC. We then discuss what would happen when these were combined together while providing some fundamental analyses on the convergence as well as clarifying the functionality requirements in each software component. We finally introduce our ongoing efforts on our software stack tool integration, which will ultimately lead to the convergence of malleability and power management, as e.g., targeted in the HPC PowerStack efforts~\cite{powerstack-community}. 



\section{Technology Trend}\label{literature}
In this section, we first summarize the trend of hardware architecture in HPC systems. 
We second introduce several prior and ongoing efforts for the malleability support in HPC systems. 
We third highlight existing co-scheduling techniques for HPC systems. 
We finally present power management studies in HPC systems.  

\subsection{Hardware Architecture}
Driven by the end of Dennard scaling in mid 2000s, the industry had to change their system designs toward multi-core and heterogeneous systems, instead of merely increasing the clock frequency~\cite{dark-silicon,moore-end}. 
As a consequence, CPU-GPU heterogeneous supercomputers have appeared around a decade ago, and now about 30\% of the HPC systems ranked in the Top500 list are equipped with GPUs~\cite{top500}. 
Nevertheless, we are now facing another serious issue, namely the slowing of Moore's law, and with that the end of the exponential growth that continued over the past 50 years is inevitable in the near future~\cite{moore-end}. 
To keep the historical performance or energy-efficiency growth ratio, both hardware-/software-level system optimizations or even radical redesigns are essential. 
To this end, adopting extremely heterogeneous architectures that consist of multiple different specialized hardware components is a promising solution, in particular to maximize the performance or energy efficiency of various common HPC workloads~\cite{extreme-hetero}. 
\textit{However, this hardware architecture trend, i.e., compute nodes will become fatter and more heterogeneous, will make it even harder to fully utilize the available resources, which will require more sophisticated resource management methodologies including co-scheduling, power management, and malleability support. }

\subsection{Malleability Support}
Malleability is the property of jobs or applications to remap themselves to varying numbers of compute resources at runtime~\cite{malleability-def}.  
When these resources are CPU cores in a shared memory environment, this kind of remapping requires less complicated data movements.  
In contrast, when whole nodes are added or removed from the resources available to a job or application, then network-based data re-distributions need to take place.  
In addition to this, communication software needs to be able to account for these changes, and update its internal data structures to represent the changes in resources.  
This is the case with  MPI libraries or PGAS run-time systems.  
There has been active research in both shared-memory~\cite{schreiber-openmp} and distributed-memory~\cite{malleability-virtual,malleability-cosh,malleability-charm,malleability-isaias,malleability-tsunami} malleable systems. 


In distributed memory systems, as may be expected, the number of changes to support malleability is larger:
Nearly the entire software stack needs to be updated to support malleability.
The scheduler, node management, process managements, communication libraries, programming models, tools and applications, among other things, require changes to support malleability in distributed memory systems.
Furthermore, existing elastic distributed memory systems, such as cloud software stacks, are incompatible with the bulk synchronous patterns that are common in scientific and engineering simulations.
Therefore, systems like Kubernetes, that already support malleability for cloud computing workloads, cannot be reused without important changes.  
In spite of the larger scope and challenges, researchers have been exploring updates to systems such as schedulers~\cite{malleability-virtual,malleability-cosh}, programming models~\cite{malleability-charm,malleability-isaias} and applications~\cite{malleability-tsunami} in the distributed-memory supercomputing field in recent years. Further, \textit{as power is becoming more and more precious in supercomputers, in particular for over-provisioned systems, we should target power budgeting and compute resources at the same time. }


\subsection{Co-scheduling}
Ever since multi-core processors appeared on the market, a variety of co-scheduling techniques have been widely studied. 
In general, these techniques are useful in order to fully utilize the resources inside of a chip/node by mixing processes/applications/jobs which require complementary resources. 
R. Cochran et al. proposed Pack \& Cap that co-locates multi-threaded applications on a multi/many-core processor while optimizing the number of threads for each of the applications~\cite{cos-cpu-powcap}. 
M. Bhadauria et al. explored the feasibility of space-shared scheduling using a greedy-based co-run job selection and resource allocation policy~\cite{cos-ics}. 
J. Breitbart et al. created a resource monitoring tool useful for co-scheduling HPC applications~\cite{cos-icppw} and provided a memory-intensity-aware co-scheduling policy~\cite{cos-cluster}. Q. Zhu et al. targeted CPU-GPU heterogeneous processors and proposed a co-scheduling approach suitable for them~\cite{cos-ipdps2}. Others examined the impact of hardware cache partitioning when co-running HPC jobs~\cite{ca-cos}. 
\textit{In general, these seminal studies are not aware of malleable HPC applications. }

\subsection{Power-Aware HPC}
Since power consumption has become the first class design constraint when building supercomputers, there have been a variety of activities or studies on power-aware HPC. 
T. R. W. Scogland et al. developed a comparative power measurement methodology through the Energy Efficient HPC Working Group, which is used for the
Green500 ranking today~\cite{eehpc}. 
T Patki et al. firstly explored the feasibility of over-provisioning for HPC systems~\cite{over-provisioning}, and following this study, there have been various resource management and scheduling researches for over-provisioned and power-constrained HPC systems~\cite{power-scheduler,power-scheduler2,power-scheduler3}. 
The PowerStack initiative community~\cite{powerstack-community} was launched based on these studies, and \textit{now we should extend the scope to cover malleability and co-scheduling to fully exploit the energy efficiency of HPC systems. }

\section{Problem Statement}\label{problem_statement}
Our ultimate goal is to provide a software stack that is capable of handling malleable jobs while providing co-scheduling and power management features for near-future over-provisioned and power-constrained HPC systems. In this section, we cover the fundamental aspects such as job classification and the relationship between malleability, co-scheduling, and power management. 

\subsection{Job Classification}
\begin{figure}[t]
\begin{center}
\begin{tabular}{c}

\begin{minipage}{0.5\hsize}
\scriptsize
\begin{center}
\makeatletter
\def\@captype{table}
\makeatother
  \begin{tabular}{ |c||c|c|c|c|}
    \hline
     & Malleability & \shortstack{ Accept \\ Slowdown} & \shortstack{Power \\ Capping} & \shortstack{Node \\ Sharing} \\\hline\hline
    \shortstack{Job \\ Class1} &  &  & & \\\hline
    \shortstack{Job \\ Class2} &  & $\checkmark$ & $\checkmark$ & $\checkmark$ \\\hline
    \shortstack{Job \\ Class3} & $\checkmark$ &  & & \\\hline
    \shortstack{Job \\ Class4} & $\checkmark$ & $\checkmark$ & $\checkmark$ & $\checkmark$ \\\hline
  \end{tabular}
  \caption{Job Classification}
\label{classification}
\end{center}
\end{minipage}

\begin{minipage}{0.5\hsize}
  \begin{center}
    \includegraphics[width=\linewidth]{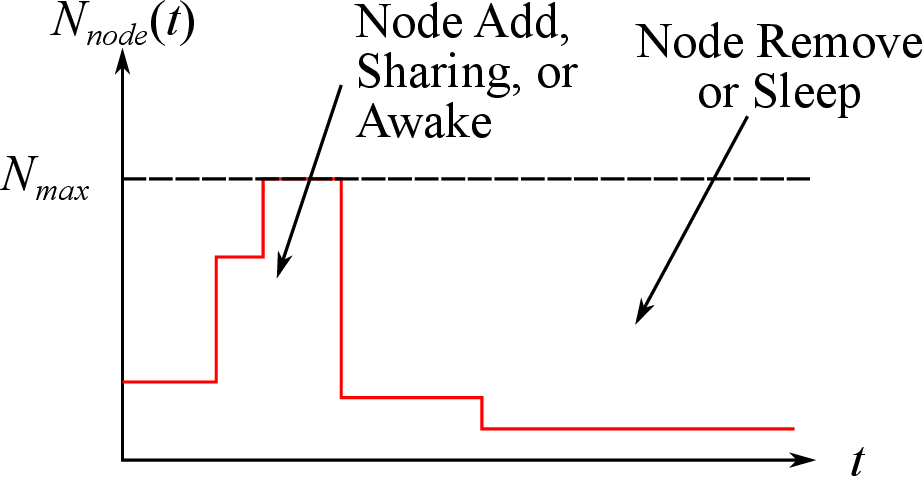}
    \caption{Malleable Job}
  \label{malleable-job}
  \end{center}
\end{minipage}

\vspace{10pt}

\\

\begin{minipage}{0.45\hsize}
  \begin{center}
    \includegraphics[width=\linewidth]{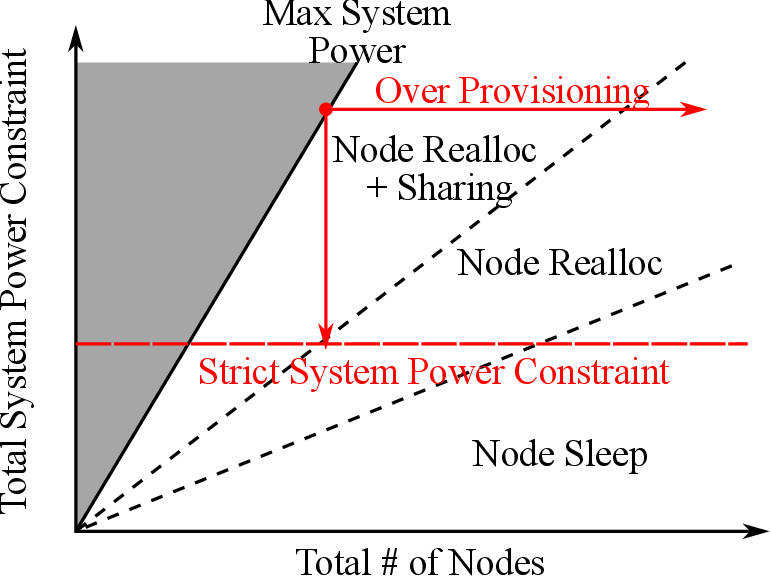}
    \caption{Malleability Handling Policies and Power Constraint}
  \label{policies}
  \end{center}
\end{minipage}

\begin{minipage}{0.55\hsize}
\scriptsize
\begin{center}
    \includegraphics[width=\linewidth]{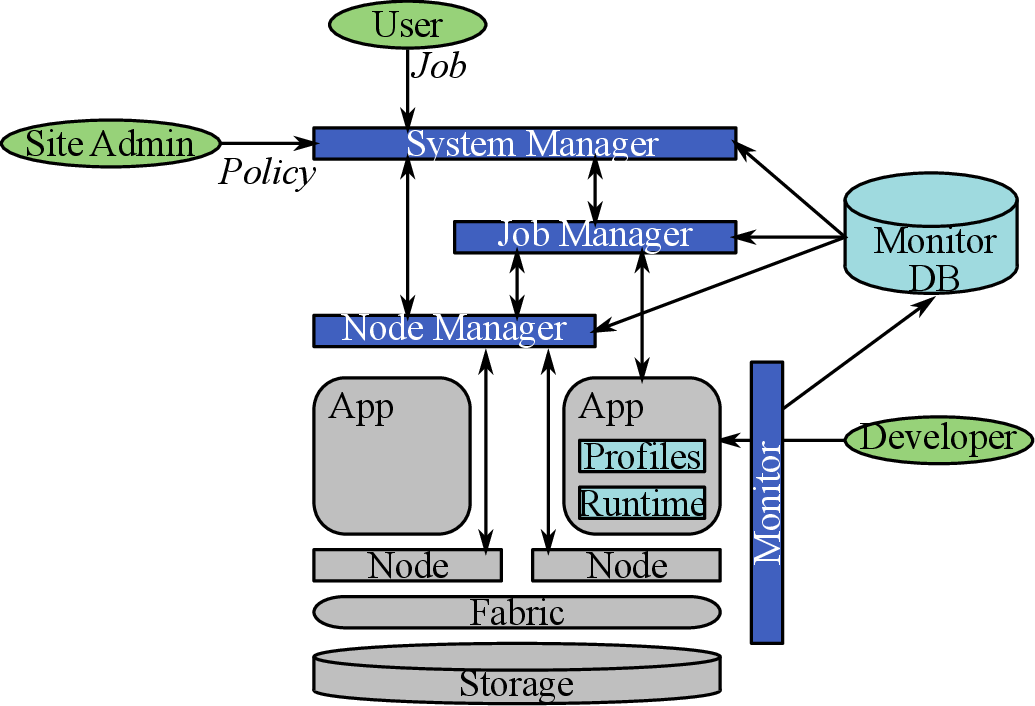}
    \caption{Our Target Strawman Architecture}
  \label{strawman}
\end{center}
\end{minipage}

\end{tabular}
\end{center}
\vspace{-5pt}
\end{figure}

Before go into the details, we first classify jobs with respect to the applicability of the advanced resource management features in Table~\ref{classification}. 
The classification is based on the following two points: (1) whether or not the application supports the malleability; (2) whether or not the user accepts the slowdown caused by the power capping and/or node sharing (or co-scheduling). 
Even though introducing the malleability feature has various advantages by exploiting the dynamic behaviors of both systems and applications, as it requires code modifications, which can be significant depending on the complexity of the application, some users may choose the traditional rigid option. 
Similarly, some users may prefer to exclusively utilize compute nodes without any power capping 
even if the administrators encourage the users to accept the slowdown (with an acceptable performance degradation rate) by offering them incentives, in terms of such as queuing priority and pricing. 
Therefore, these different classes of jobs will co-exist in future HPC systems with these advanced resource management features, and the entire software stack as well as the administrators must carefully handle them accordingly, in terms of resource allocations, queuing priority, token accounting, and so forth, which will be discussed later in this paper.


\subsection{Malleable Jobs under Power Constraint}
Next we focus on malleable jobs and their dynamic behaviors (\textit{Job Class 3/4}) as depicted in Figure~\ref{malleable-job}. 
The X-axis indicates time ($t$), while the Y-axis represents the number of requested nodes or the scale of MPI rank ($N_{node}(t)$). 
$N_{max}$ is the maximum of $N_{node}(t)$ throughout the job execution. 

We have several options to deal with the dynamic resource re-allocations to malleable jobs, and an optimal choice highly depends on the remaining resources in terms of both compute nodes and power. 
If available compute nodes are not very plentiful, dealing compute nodes across the job is a suitable choice. 
If no compute node is available (but power budget is still remaining), node sharing (or co-scheduling) should be considered. Note that the system must care about the job classes and the acceptable slowdown ratios designated by the users in this case. 
For an over-provisioned system with a very strict power constraint, allocating $N_{max}$ nodes to a malleable job and sleeping/awaking them to virtually deal compute nodes across different jobs will be an option to be considered. 
As available compute nodes are very plentiful (but most of them must be in the sleep state) in such a system, the malleability can be handled by just turning them into the sleep/awake state. 
Furthermore, co-optimizing both the node and power allocations to applications in order to maximize performance under a power cap would be another promising option. 

Figure~\ref{policies} intuitively summarizes the conditions mentioned above, and it is important for current/future HPC systems to analyze, model, and quantify the the exact boundaries to determine the policy selection from these different resource management options. 
This exploration is a new research opportunity, and we need theoretical studies to demystify them by using such as job traces obtained from supercomputers and putting them into simulators with realistic setups. 
In this fundamental study, we will estimate or set some assumptions to classify the jobs in the trace into the categories shown in Table~\ref{classification}, in terms of number of jobs, job scale distributions, and execution time distributions, because the optimal policy setups and the effectiveness of these approaches will highly depend on these factors. 
To make the exploration more realistic, analyzing the scalability of representative applications at the granularity of application phase will help to assess the dynamic behavior. 
Further, quantifying the benefits of our approach from both the systems' and users' point of views will be essential, which  includes the exploration on what incentives we should provide to users. 
Once a policy selection methodology established throughout this study, that will be deployed/implemented on the software stack.

\section{Toward Convergence of Malleability and PowerStack}\label{solution}
Driven by the problem statement and the basic assessment described in the last section, here we describe our high-level solution and ongoing efforts to realize it. 
First, we introduce our reference strawman software architecture. 
Second, we explain our high-level architectural solution and the detailed roles of components in the strawman architecture. 
We then finally highlight our ongoing efforts on the software integration to realize it. 

\subsection{Strawman Architecture}\label{strawman_arch}
Figure~\ref{strawman} illustrates our high-level software architecture, consisting of several components and actors. 
The roles of the components are summarized as follows: 

\vspace{-10pt}

\subsubsection{System Manager} 
The system manager receives a set of jobs to be scheduled within the system and indicatively decides upon when to schedule each job, to which specific compute nodes to map it, and under which power budget or setting. It also handles any dynamic resource/power requests from the job/node managers at runtime. 

\vspace{-10pt}

\subsubsection{Job Manager}

The job manager performs optimizations considering the performance behaviour of each application, its fine-grained resource footprint, its phases and any interactions/dependencies dictated by the entire workflow. 
It provides an option to users for a fine-tuned application-level hardware knob controlling and also provides the functionality to scaling up/down the job size.  


\vspace{-10pt}

\subsubsection{Node Manager}
The node manager provides access to node-level hardware controls and monitors. Moreover, the node manager implements processor level and node level power management or resource partitioning policies, and it mediate all the hardware control requests coming from the software stack. 

\vspace{-10pt}

\subsubsection{Monitor}
The monitor is responsible for collecting in-band and out-of-band data for performance, scheduling, resource management, and so forth. The monitor operates continuously without interfering with running jobs, and collects, aggregates, records, and analyses various metrics, and pushes necessary real-time or profiling data to the other components.

\vspace{-10pt}

\begin{figure}[t]
  \begin{center}
    \includegraphics[width=0.95\linewidth]{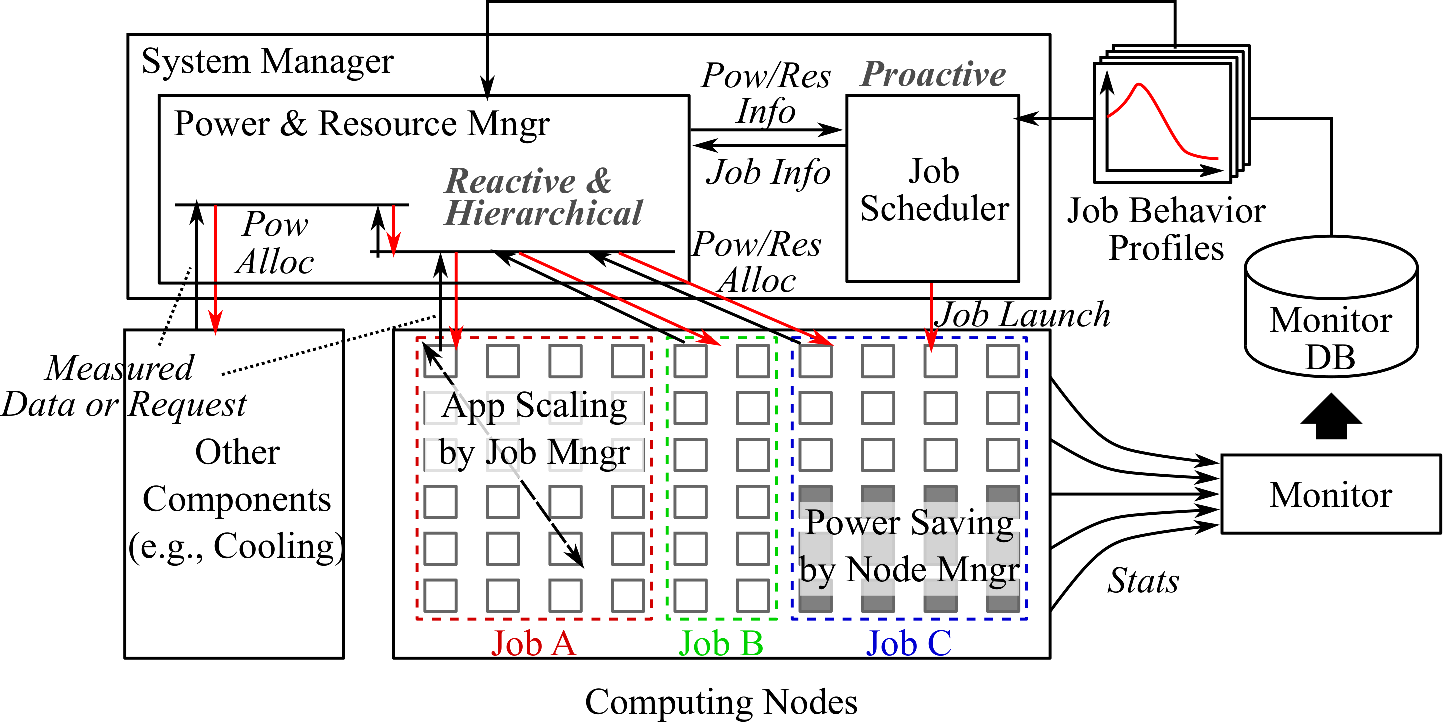}
    \caption{Hierarchical and Dynamic Resource Management}
  \label{concept}
  \end{center}
  \vspace{-10pt}
\end{figure}

\subsection{Solution Overview and Requirements}
Figure~\ref{concept} illustrates the high-level concept of our solution. Overall, we apply a hierarchical and feedback-driven resource management approach. The requirements for each software component and the administrators are as follows:

\subsubsection{System Manager}
The system manager mainly deals with two different tasks: (1) job scheduling, which is based on proactive decisions; and (2) dynamic and reactive resource adjustment across jobs/nodes. 
For the former, as the job scheduling decisions cannot be changed after job launches (unless we apply check-pointing and migrations that however can induce significant overheads), they are basically proactive relying on static information such as job profiles. 
Naive heuristics such as the FCFS with back-filling are widely used, however as we support malleable jobs, prediction-based approaches will be more important, e.g., those based on estimated dynamic behaviors of malleable jobs using such as profiling or any other information given by such as users. Further, as we apply co-scheduling and/or power capping to some classes of jobs, we need to revisit even the conventional back-filling strategy, e.g., choosing back-filling jobs based also on the remaining/requested power budget and the impact of node sharing. 
As for the dynamic resource management, we consider a hierarchical approach as shown in the figure: (1) jobs trade the power budgets and nodes by interacting with the system manager depending on their needs; (2) the system manager monitors the remaining nodes/power and governs the redistribution based on the requests; and (3) the node manager scales the power cap accordingly and optimizes the node resource partitioning (if co-scheduled). 
Note that if other components (e.g., I/O nodes or cooling facilities) support the power capping capability, the system manager should handle the power trading across compute nodes and them as well. 

\subsubsection{Job Manager}
The major role of the job manager in this software stack is supporting the malleability functions and providing proper interfaces to the system manager, the node manager, and the developers. 
Beyond that, as the power budgeting should be supported for over-provisioned and power-constrained systems, the interface should also be able to handle the power budget requesting functionalities, not limited to removing/adding compute nodes. 
The power management should cover not only the sleeping/awaking decisions for the malleability behavior, but also should care about the per-phase/-loop characteristics (e.g., compute intensity, cache hit/miss, accelerator utilization, etc.) to determine the setups of hardware knobs during the active state. 
The characterizations should be based on such as profiling provided by the monitor tool, and the hardware knob setups are sent to the node manager and are handled by it. 

\subsubsection{Node Manager}
The node manager mainly focuses on the hardware knob control, instructed by the job manager, or optimize the knobs by itself if the job manager (or the application) doesn't apply any application-oriented optimizations. 
It optimizes the power knob setup for each target hardware component inside a node while keeping the total node power constraint given by the system manager (or the job manager). 
In case the node becomes idle and unused for a malleable job, that should turn the node into the sleep state based on the instruction given by the system/job manager. 
Further, if multiple jobs are running the same node in a space sharing manner, it should handle the resource partitioning properly to meet the performance requirement for both of the co-running applications. For these decisions, the node manager can utilize statistics given by the monitor tool. 

\subsubsection{Monitor}
The monitoring tool must be able to keep track of the power, resource usage information, and so forth, associated with each job/application, which will be ultimately utilized for various objectives. 
For instance, these collected information is useful for the other software components in their decisions, mainly the job profiling purpose as described before. 
In addition, the collected power/resource utilization information should be used for the human actors. 
One example is the job pricing purpose determined by the site administrators, which will be described later. 
Further, more advanced functionalities including modeling and analysis would help. 
One option is pointing out resource wastes or potential benefits of introducing malleability, power capping, or co-scheduling for users who submit jobs belong to \textit{Job Class 1} (see Table~\ref{classification}). 
More specifically, notifying the estimated queuing time and cost reduction by applying/accepting these features would be a great encouragement for users to apply/accept these features.

\subsubsection{Site Administrators}
One of the major roles of the administrators is setting up the system configurations, including the total system power constraint or dividing the job queue per job class. 
Further, as introducing malleability into an application requires extra efforts to modify their codes, the administrators need to clarify the benefits to encourage their use. 
This is also the case for applying power capping and/or co-scheduling as they incur performance degradation even though the resource manager attempts to minimize the impact. 
One option is taking these advanced resource management features into account in the token consumption calculations, i.e., how much they charge for a job. 
The cost is usually calculated based on the number of occupied nodes multiplied by the job runtime. As the number of nodes dynamically changes during the execution of a malleable job, the cost should be significantly reduced. Further, if the power capping and/or co-scheduling is applied to the job as well, that should be also reflected on the cost, i.e., energy-based pricing or interference-aware pricing~\cite{pricing}.

\subsection{Our Ongoing Efforts on Software Tool Integration}

To realize the high-level solution described above, several software integration projects are ongoing. 
One is DEEP-SEA project~\cite{deep-sea} that aims at providing a programming environment for European exascale systems, which includes the malleability support, and the other one is REGALE project~\cite{regale} that focuses on realizing the HPC PowerStack~\cite{powerstack-community}, including both power management and co-scheduling aspects. 
In this subsection, we briefly introduce the current status of them, and our ambition is combining these two integration paths together in the near future. 

\subsubsection{Malleability Support}

At the system management level, we are engaged in experimental development with resource managers like FLUX~\cite{flux} and Slurm~\cite{slurm} as a part of DEEP-SEA project~\cite{deep-sea}.  
These are being extended with dynamic job allocation functionalities.  
In addition to this, new experimental scheduling heuristics are being developed. 
Currently, these are extensions to the well established FCFS with back-filling heuristic already available in these frameworks.  
We are also in collaboration with developers of monitoring and data-capture frameworks, such as DCDB~\cite{dcdb}, to identify metrics that are relevant to allocation size updates in jobs.
These monitors are being updated to capture data of jobs with changing allocation sizes.

A set of application processes is created for workloads to run in our systems, in one or more nodes.
These processes require relevant metadata and synchronization operations to establish communication, among other things.
Between these system managers and the run-time systems of programming models, there are process management interfaces, that allow the exchange of such metadata.
These interfaces have traditionally been vendor specific, and as a result has increased the challenge of developing run-time systems, especially in distributed memory systems.
The PMIx~\cite{pmix} standard aims to remove these additional compatibility challenges.
We are engaged in its standardization efforts, as well as in the development of its Open PMIx library.
Both the standard and the library are being extended to better support the dynamic exchange of allocation metadata, required by malleable systems.

\subsubsection{PowerStack Support}
As an initial step, we are developing a software stack to realize the PowerStack~\cite{powerstack-community} that enables a variety of power management functionalities from naive to a more sophisticated one in REGALE project~\cite{regale}. 
We have already completed defining the initial software architecture, requirements, and supported use cases, and now we are working on the software tool integration based on the architectural definition. 
For the system manager, in particular the job scheduling, we are going to cover Slurm~\cite{slurm} and OAR~\cite{oar}, i.e., we are going to provide multiple different software stack instances to realize the architecture. 
For the job manager, EAR~\cite{ear}, Countdown~\cite{countdown}, or BDPO~\cite{beo} will be used. Note EAR has a variety of functionalities ranging from the system, job, and node manager, and thus is one of the key tools. Countdown tries to minimize the power consumption while waiting for the completion of an MPI communication, by scaling down the clock frequency or going into one of the CPU sleep states (C-state). BDPO is a job-oriented profile-based power-performance optimization tool, which optimizes clock frequency to trade-off performance and energy or to minimize energy while using its phase detection mechanism. 
As for the node manager, aside from EAR, BEO~\cite{beo} and PULP Controller~\cite{pulp} are promising tools. BEO is an out-of-band power monitoring and controlling tool. PULP Controller is a low-level power controller, works transparently to the application, user, and system software, currently targeting EPI processors. 
As for the monitor, DCDB~\cite{dcdb} and EXAMON~\cite{examon} will be used, both of which support in-band/out-of-band monitoring properties as well as the functionalities to analyze/model the monitored data. 

Some of the tools have been already integrated, and other integration is under construction. 
In addition to the tool integration, some sophistication paths are ongoing, one of which is the co-scheduling support. 
After completing the integration in this PowerStack implementation phase, we are planning to extend it to converge with the malleability software stack. 


\section{Conclusion}\label{conclusion}
Near-future HPC systems will be over-provisioned and power-constrained. 
In this position paper, we explicitly target such systems and discussed the necessity/requirements of sophisticated resource handling mechanisms, i.e., the combination of malleability support, co-scheduling, and power management. 
More specifically, we first introduced the trends of HPC architectures and the prior studies on these resource management concepts. 
We second discussed what would happen when these were combined together while providing several prominent use cases as well as some fundamental analyses. We finally introduced our ongoing efforts on our software stack tool integration, which will ultimately lead to the convergence of malleability and PowerStack, leaving a significant impact on both of these communities.

\subsubsection{Acknowledgements} 
We would like to express our sincere gratitude to the anonymous reviewers for their constructive suggestions. 
This work has received funding under the European Commission's EuroHPC and H2020 programmes under grant agreement no. 955606 and no. 956560.

%
%
%
\bibliographystyle{splncs04}
\bibliography{ref.bib}

\end{document}